\documentclass[preprint,groupedaddress,footinbib,amssymb]{revtex4}

\usepackage{graphicx}% Include figure files
\usepackage{amssymb}
\usepackage{amstext}
\begin{document}

\title{Minimal Radius of Magnetic Skyrmions: \newline Statics and Dynamics}

\author{A. Siemens$^{1}$, Y. Zhang$^{1}$, J. Hagemeister$^{1}$, E. Y.\,Vedmedenko$^{1}$, R.\,Wiesendanger$^{1}$}
%\ead{vedmeden@physnet.uni-hamburg.de}
\address{$^{1}$University of Hamburg, Department of Physics, Jungiusstr. 11a, 20355 Hamburg}

%\address{IOP Publishing, DiracHouse, Temple Back, Bristol BS1 6BE, UK}
%\ead{custserv@iop.org}
\begin{abstract}
In a broad range of applied magnetic fields and material parameters isolated magnetic skyrmions condense into skyrmion lattices. While the geometry of isolated skyrmions and their lattice counterparts strongly depend on field and Dzyaloshinski-Moriya interaction, this issue has not been adequately addressed in previous studies. Meanwhile, this information is extremely important for applications, because the skyrmion size and the interskyrmion distance have to be tuned for skyrmion based memory and logic devices. In this investigation we elucidate the size and density-dependent phase diagram showing traditional phases in field vs. material parameters space by means of Monte-Carlo simulations on a discrete lattice. The obtained diagram permits us to establish that, in contrast to the continuum limit, skyrmions on a discrete lattice cannot be smaller than some critical size and have a very specific shape. These minimal skyrmions correspond to the micromagnetic configuration at the energy barrier between the ferromagnetic and the skyrmionic states. Furthermore, we use atomistic Landau-Lifshitz-Gilbert simulations to study  dynamics of the skyrmion annihilation. It is shown that this procees consists of two stages: the continuous skyrmion contraction and its discontinuous annihilation. The detailed analysis of this dynamical process is given.

\end{abstract}

%\pacs{67.85.Hj, 32.10.Dk, 75.45.+j, 75.10.-b}
\maketitle

\section{Introduction}
Chiral skyrmions are of great interest in fundamental physics and mathematics because of their non-trivial topological properties \cite{Bogdanov:J.Phys.,Manton}. Theoretically, the existence of magnetic skyrmions has been predicted in a variety of three- and two-dimensional systems \cite{Roessler:Nature,Bogdanov89,Rosza}. Recently, they have indeed been found in cubic helimagnets \cite{Boeni,Tokura,Monchesky} and in two-dimensional magnetic layers on metal substrates exhibiting strong spin-orbit coupling \cite{VonB}. In both cases skyrmions
easily condense into lattices, while isolated skyrmions are very often unstable.
For information storage applications, however, isolated magnetic skyrmions are required, as they have to be created, deleted and manipulated independently of their neighbors \cite{Fert}. In a few systems only (e.g. Fe/Pd/Ir(111)) isolated skyrmions could be created and stabilized experimentally \cite{Romming:Science,Romming:PRL2015}.

While the dependence of the size of these isolated objects on an external magnetic field and material parameters have been studied in the framework of continuous model systems \cite{Bogdanov:94,Butenko}, the experimentally found nanoscale skyrmions interact with a discrete atomic lattice. Hence, in search for stable isolated skyrmions it is important to understand whether their properties differ in continuous and discrete systems \cite{Vedmedenko:PRL2014}. Two aspects appear to be particularly important: First, according to continuum field theory skyrmions survive in infinitely large fields \cite{Bogdanov:94,Leonov}. This property is known under the name of topological protection. Experimental results, however, show that at some critical field a skyrmion on a discrete lattice will collapse \cite{Romming:PRL2015,Leonov}. Hence, it is important to study the critical field of the collapse and the skyrmion size as a function of field and other parameters. The second aspect concerns the transition between the regime of skyrmion lattices and that of isolated skyrmions. From field theory it is known that the radii of skyrmions in the two phases have a different dependence on field and Dzyaloshinskii-Moriya (DM) interaction \cite{Bogdanov89,Butenko,Leonov}. Particularly, the radius of an isolated skyrmion should increase with the strength of the DM interaction, while the radius of skyrmions in a lattice should, in contrast, decrease \cite{Bogdanov89,Butenko,Leonov}.  In the published phase diagrams \cite{Banerjee:PRX2014,Buhrandt:PRB2013,Han, Li}, however, a differentiation between those regimes is not addressed.

To close this lack of knowledge we investigate the skyrmion radius $R$ and the skyrmion density $\rho$ as a function of the ratio between the strength of the DM interaction $D$ and the exchange interaction $J$ for different fields $B$ by means of Monte-Carlo simulations. We find that the skyrmion radius strongly depends on the field and the DM interaction. In agreement with continuum model studies \cite{Butenko} the dependence $R=f(D/J)$ is different in the regimes of diluted and condensed skyrmions. In contrast to continuous systems skyrmions on a discrete lattice cannot shrink below a critical radius $R_{\rm min}$. To find this radius the numerical data have been plotted in phase diagrams $R=f(D/J, B/J)$ and $\rho=f(D/J, B/J)$. We find that the smallest skyrmions occur in the diluted regime and that $R_{\rm min}$ itself is $B$ and $D/J$ dependent. We use this knowledge to study the dynamics of the skyrmion collapse by means of atomistic Landau-Lifshitz-Gilbert simulations (LLG). It is found that the ultimately small skyrmion has a finite diameter of approximately two lattice constants and a very specific shape. Its energy defines the energy barrier between the skyrmionic and the ferromagnetic states. The process of skyrmion erasing consists of two phases: the longer period of the skyrmion size minimization and the abrupt skyrmion annihilation.

\begin{figure}
\includegraphics*[width=1.0\columnwidth]{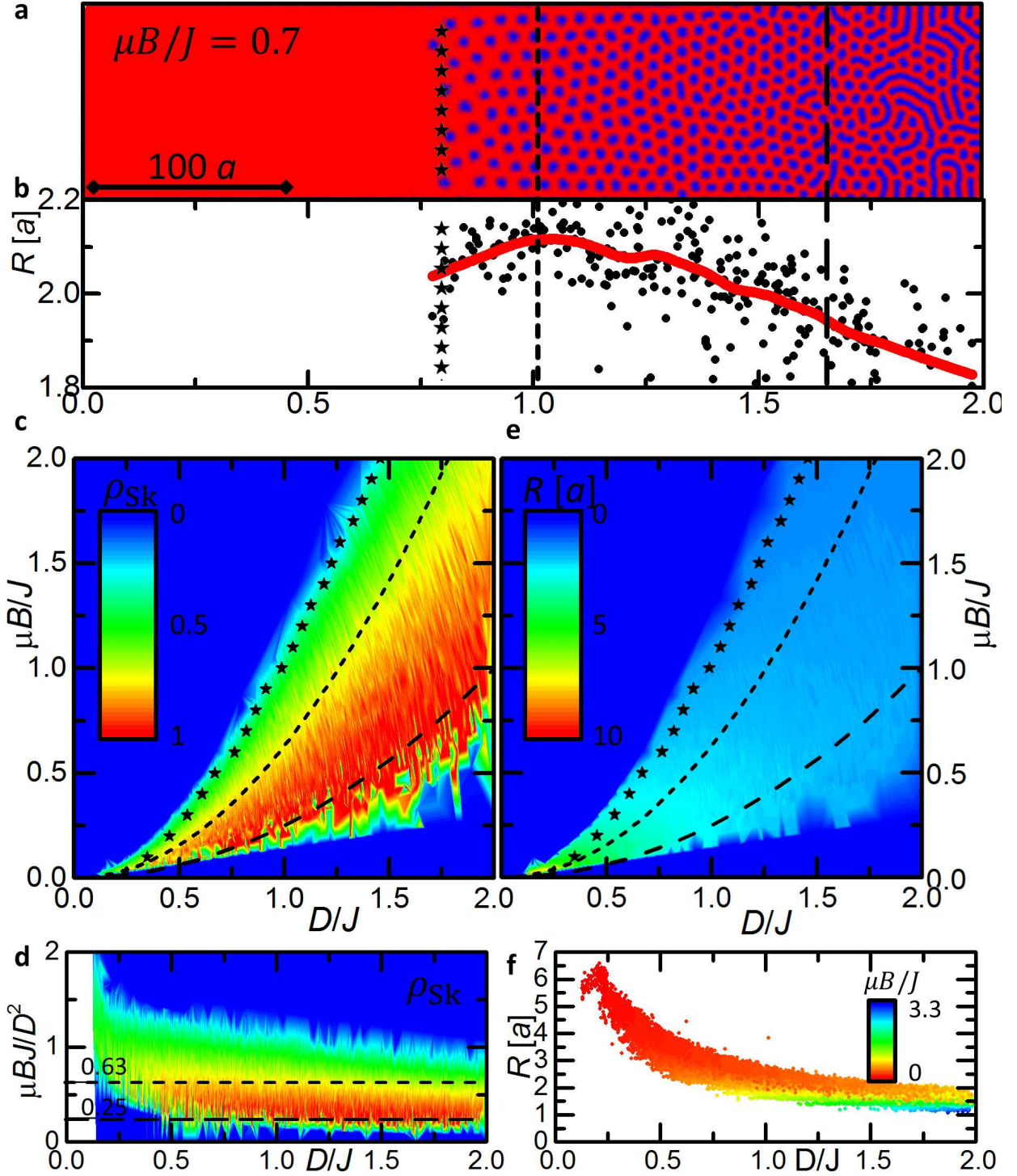}
	\caption{
	(a) Equilibrium micromagnetic configuration obtained in MC simulations at $\mu B/J = 0.7$ and $k_\mathrm{B}T/J = 0.1\,J$ with a gradient of the DM interaction strength along the $x$-axis enforcing a transition from the FM state to the spiral state. (b) Visualization of the corresponding skyrmion radii. Points represent the numerical data, while the solid line is the averaged $R$ value. (c,d) Phase diagrams using the skyrmion density $\rho$ as an order parameter. (e,f) The skyrmion radii as a function of the position in phase space.
	}
	\label{fig:mp}
\end{figure}

\section{Simulation Methods}

The skyrmionic systems have been calculated on a two-dimensional hexagonal
lattice corresponding to the hcp(0001) or fcc(111) crystallographic surfaces. At every lattice point a three dimensional (Heisenberg) magnetic moment is assumed. For the sake of generality, an ultrathin magnetic film is described by the standard effective Hamiltonian characteristic for skyrmionic systems:
\begin{equation}
H = -J \sum_{<i,j>}\textbf{S}_i \cdot \textbf{S}_j - \sum_{<i,j>} \textbf{D}_{i,j} \cdot \left( \textbf{S}_i \times \textbf{S}_j \right)
- \mu\sum_i{\textbf{B} \cdot \textbf{S}_i} - K \sum_i S_{i,\mathrm{z}}^2
\label{eqn:hamilton}
\end{equation}
where $\mathbf{S}_{i}$ is a three-dimensional unit vector, $\mu$ the localized atomic magnetic moment, $\textbf{B}$ an external magnetic field, $K$ a magnetic anisotropy, $D = |\textbf{D}_{i,j}|$ the strength of the Dzyaloshinskii-Moriya interaction and $J$ is the exchange constant. Systems consisting of up to 300.000 sites using open boundary conditions have been investigated in MC simulations, while systems up to 1.000 sites with open or fixed boundaries were used in LLG spin dynamics.

At each Monte-Carlo step the magnetization of a randomly chosen spin has been sampled according to the Boltzmann probability distribution. In order to achieve an equilibrium state a standard  slow tempered annealing has been applied. Each run consisted of up to $N_t=75\cdot 10^3$ MC steps per temperature with 20 temperature steps. The new configuration has been accepted or rejected with a single spin update based on the Metropolis algorithm. For the statistical evaluation several runs with different seed numbers have been analyzed.

For an effective exploration of the phase space standard calculations have been combined with the calculations of very large samples with spatial gradients of the $D/J$ ratio. Several gradient values have been explored and compared with calculations using constant simulation parameters. It has been found that for $d(D/J)/dr\lesssim 0.01/a$, with $a$ the lattice constant, the two methods provide identical results. We typically use $d(D/J)/dr=5\cdot 10^{-3}/a$ in the simulations. The calculations using the interaction gradients, however, are less time consuming and provide the possibility of direct visualization and analysis of the magnetic microstructure \cite{Vedmedenko:JMMM2003,Millev:JPhysD2003} (see an example in Figs.1,2). The analysis of the skyrmion geometry as a function of the parameter $D/J$ yields with important information on the phase boundaries as will be shown below.

For the spin dynamics calculations \cite{Stapelfeldt} the motion of each dipole on a lattice has been described by the generalized Landau-Lifshitz-Gilbert equation at zero temperature:
\begin{eqnarray}
    \label{LLG}
    \frac{\partial {\vec{S}}_i}{\partial t} = &-&
    \frac{\gamma}{\left(1+\alpha_D^2\right)\mu_S}{\vec{S}}_i \times \left[
    {\vec{ H}}_i + \alpha_D \left({\vec{ S}}_i \times {\vec{H}}_i
    \right)\right]\;,
 \end{eqnarray}
with the gyromagnetic ratio $\gamma$, the Gilbert damping $\alpha_D=0.01$, and the internal field $\vec H_i=-\partial \mathcal H/\partial \vec S_i$. The first term describes the precession movement of the spin around the internal field vector while the second term describes a circular motion towards $\vec H_i$. The simulation parameters were chosen to match the material Pd/Fe/Ir(111) for which skyrmions already have been studied \cite{Romming:Science}. The chosen parameters were taken from recent calculations \cite{Hagemeister}. We set the strength of the interactions at $J= 7.0$ meV and $D= 2.2$ meV. The magnetic moment was set to $\mu \approx 3 \mu_B \approx 0.2$ meV/T.

\section{Phase Diagram}

In phase diagrams of skyrmionic systems with an interface induced DM interaction, three main phases are typically distinguished from each other including the skyrmion lattice phase (Skx), the spin spiral phase (SS) and the ferromagnetic phase (FM) \cite{Banerjee:PRX2014,Han,Duine:PRB2015}. Those phases are studied as a function of either temperature or interaction strength. In both cases the winding number Q is often used as the order parameter. The winding number doesn't provide any specific information about the skyrmion size or about the density of a skyrmion lattice but rather defines the very existence of these topological objects. Hence, two skyrmion lattices of different density and different skyrmion sizes but with identical numbers of skyrmions have identical winding numbers $Q$ and, hence, are indistinguishable in common phase diagrams. It is known, however, that the skyrmion size is strongly parameter-dependent \cite{Romming:PRL2015,Butenko}.

The aim of this investigation is the derivation of equilibrium states of chiral systems as a function of the effective interaction strength including the information on the skyrmion size and density. For that purpose, similarly to \cite{Tokura}, we define the density order parameter $\rho$ as the ratio between the effective area occupied by skyrmions to the area occupied by a closed packed lattice of skyrmions having the same radius $R$.
The skyrmion radius $R$ has been determined numerically by a two-dimensional surface approximation of the azimuthal angle $\Theta$ of the local magnetization texture with a formula which was used originally to describe $360^\circ$ N\'eel domain walls \cite{Kubetzka}. In this approximation the angle of the local magnetic moment is given by
\begin{equation}\label{eq:phi}
    \Theta_{360}(r)= \arcsin\left[\tanh\left(\frac{r+c}{w/2}\right)\right] + \arcsin\left[\tanh\left(\frac{r-c}{w/2}\right)\right]
\end{equation}
defining a full rotation of $2\pi$ from one to the other side of the domain wall with the fit parameters $c$ and $w$, and r the distance from the center. According to the previous investigation \cite{Romming:PRL2015} the magnetization profile of a skyrmion can be well approximated by this formula. The skyrmion diameter can then be determined as the size of the region in which the magnetization has a component in the direction opposite to the magnetic field.

To obtain the phase diagrams very large samples ($3\times 10^5$ sites) of rectangular shape with slow $D/J$ gradient along the $x$ axis have been equilibrated at various magnetic fields $B$. The magnetocrystalline anisotropy was set to zero for simplicity. The obtained results have been cross-checked via standard calculations without any gradient as explained in the previous section. Data presented here correspond to $\delta (\frac{D}{J})/\delta r\approx 0.005/a$ or $ {\delta D}/{\delta r}\approx $ 0.035 meV/$a$. The method of gradients permits the direct visualization of the magnetic microstructure as a function of the strength of DM interaction. Fig.1a shows the equilibrium micromagnetic structure of a sample subject to a magnetic field of strength $\mu B/J = 0.7$ at $k_\mathrm{B}T/J = 0.1\,J$ in which the transitions from the FM to the Skx, and eventually to the SS phase driven by an increasing DM-interaction can be observed.
Additionally to this information one immediately recognizes a variation of the skyrmion size and density with increasing $D/J$. Fig.1b exhibits the corresponding $D/J$ dependence of the skyrmion radius. Here, the numerical values are represented by dots, while the mean skyrmion radius $R$ is given by the solid red line.

Interestingly, one can observe a drastic change in the behavior of the skyrmion radius at $D/J\approx 1.05$. According to considerations using a continuum model for the description of skyrmions, the skyrmion radius should be $R\propto D/B$ for isolated skyrmions but follows the opposite $R \propto J/D$ dependence in skyrmion lattices \cite{Butenko}. We can confirm this behavior for skyrmions on discrete lattices and use it as a suitable parameter to identify the position of the transition from the FM phase with occasional isolated metastable skyrmions to the dense skyrmion lattice by the maximum $R$ value. The resulting phase diagrams are shown in Fig.1c,d. In both cases the phase diagram is colored using the density order parameter $\rho$, but two different scaling schemes ($\rho=f(D/J, B/J)$ and $\rho=f(D/J, \mu BJ/D^2)$) are used. The short-dashed lines in all panels of Fig.1 indicate the transitions between the phase of isolated skyrmions and that of the skyrmion lattice, while the long-dashed lines represent transitions between the Skx and the SS phase. Both dashed lines correspond to the functions of the type $\mu B_\mathrm{A|B} J/D^2$ first defined in \cite{Banerjee:PRX2014,Duine:PRB2015}, where $B_\mathrm{A|B}$ gives the critical field of the transition between phases A and B.
In phase diagrams of Fig.1, ${\mu B^\mathrm{FM|Skx}J}/{D^2} \approx 0.63$ and ${\mu B^\mathrm{Skx|SS}J}/{D^2} \approx 0.25$ have been found to separate the corresponding phases. While in \cite{Banerjee:PRX2014,Duine:PRB2015} no distinction between the Skx and the isolated skyrmions has been made, in our calculations at finite temperatures the isolated skyrmions survive up to the boundary indicated by the stars. The transition from the skyrmion lattice to the spin-spiral state appears to be diffuse: a rather broad region exists in which skyrmions and elongated spiral-like textures coexist as can also be observed in Fig.1a.

Our calculations show that the density order parameter permits to identify the phase of isolated skyrmions additionally to the commonly addressed Skx and SS phases. The phase diagrams presented above are in good agreement with investigations \cite{Banerjee:PRX2014,Duine:PRB2015}, which show that functions $\mu B_\mathrm{A|B} J/D^2$ should be constant at the FM|Skx and Skx|SS phase boundaries. While we confirm the linearity of the functions presented above at $D/J>0.5$, we find some non-linearities at weaker $D/J$. The reason seems to lie in the interplay between the increasingly large skyrmion radius in the isolated skyrmion phase and the finite size of the sample. The strong $D/J$ dependence of the skyrmion radius can be appreciated in Fig.1e,f. Furthermore, the skyrmion radius decreases with increasing magnetic field for a given set of $D/J$ (Fig.1f) and vanishes at a finite minimal size defined by the bottom borderline of Fig.1f.

\begin{figure}
\includegraphics*[width=1.0\columnwidth]{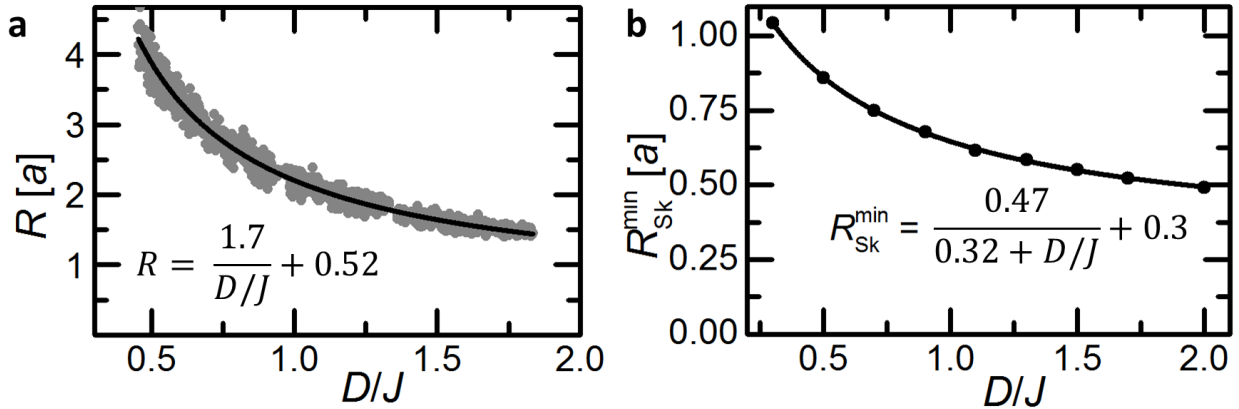}
	\caption{(a) The equilibrium skyrmion radius at the phase boundary separating the Skx and FM phases obtained in the MC simulations at $k_\mathrm{B}T/J = 0.1\,J$. (b) The dynamical critical skyrmion radius derived from LLG simulations at zero temperature. For smaller radii, the system inevitably relaxes to the FM state.
	}
	\label{fig:F2}
\end{figure}

The last observation is particularly interesting, because it is in contrast to calculations using a continuous description of the skyrmions \cite{Bogdanov89}. The skyrmions in continuum can have any non-vanishing size due to the topological protection that doesn't allow for transitions between the skyrmionic and a field polarized state. In discrete systems, the topological protection rather manifests itself in the combination of a finite energy barrier and the attempt frequency \cite{Hagemeister}. Fig.1f indicates that the minimal skyrmion size decreases with an increasing ratio of $D/J$ which is reasonable since the DM interaction stabilizes the skyrmionic magnetic texture.

We start a more detailed analysis of this issue by investigating the skyrmion radius at the phase boundary between the skyrmion lattice and the diluted phase at $\mu B^\mathrm{FM|Skx}J/D^2 \approx 0.63$ given in Fig.2a. Indeed, the skyrmion radius decreases with an increasing $D/J$ ratio and may be well approximated by a hyperbolic function with an offset of $(0.525\pm 0.008)\,a$ which gives the smallest possible radius for a skyrmion at the phase boundary at large $D/J$. However, metastable skyrmions can have extended lifetimes well beyond this phase boundary in dependence of the temperature. Especially at zero temperature, a skyrmion will possess an infinite lifetime as long as it rests in a local energy minimum protected by a finite energy barrier against a transition to the energetically lower field polarized state. Therefore, the interesting question is, at which critical skyrmion size does the system overcome the separating energy barrier and inevitably relaxes into the ferromagnetic state. This issue was studied by means of LLG simulations at zero temperature. The results are presented in Fig.2b. At this point we have to mention that the MC simulations of Fig.2a concern the equilibrium skyrmion sizes, while the dynamical simulations of Fig.2b show the radii of unstable skyrmions at the very moment, when the energy barrier disappears. Therefore dynamical skyrmion radii are smaller than those of Fig.2a. However, the statical and dynamical $D/J$ dependencies have the same functionality. The minimal skyrmion radii lie in the region of $(0.5-1)\,a$ for the investigated parameter range of $D/J \in [0.3,2.0]$. In the following the dynamics of the skyrmion annihilation is studied in more detail by means of LLG simulations.

\begin{figure}
\includegraphics*[width=1.0\columnwidth]{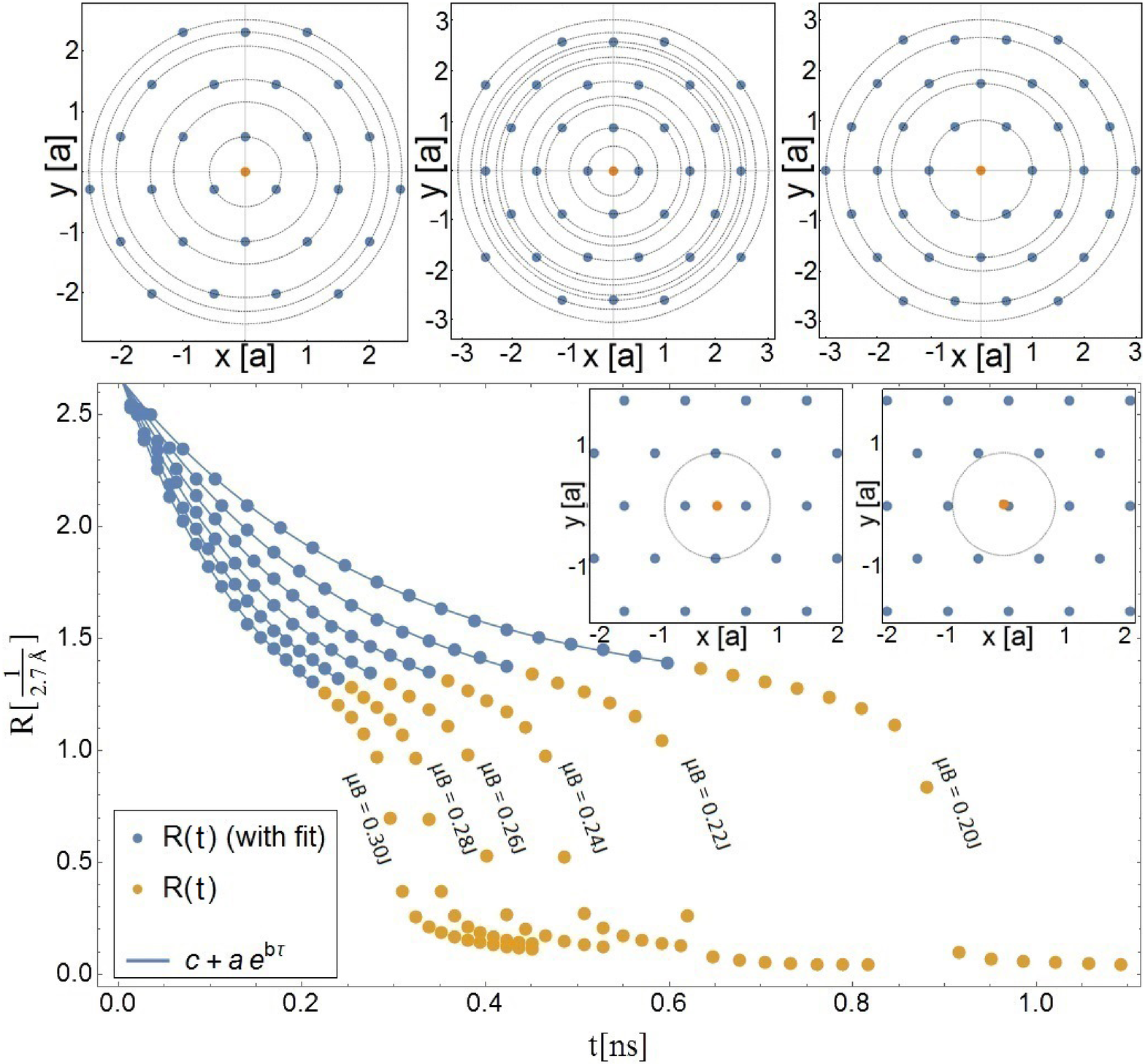}
	\caption{Upper panels show schematics of three possible rotationally symmetric configurations of a smallest possible skyrmion on a hexagonal lattice. Bottom panel shows the numerically calculated time evolution (points) and fits (solid lines) of the skyrmion radius at different applied magnetic fields $B$ at zero temperature. The part of numerical data used for fits is darker. Field values are expressed in units of the exchange interaction constant $J$. The insets in the bottom panel show smallest skyrmions obtained in LLG spin dynamics.
	}
	\label{fig:F3}
\end{figure}

\section{Deleting Skyrmions}

Since future magnetic data storage is expected to use isolated skyrmions, we now focus on this kind of skyrmions only. In the following we use the energy parameters and atomic lattice structure characteristic for the Pd/Fe/Ir(111) system; i.e., $J\approx 7$ meV and $D\approx 2.2$ meV. The size of the investigated system was adapted in such a way that only one single skyrmion is being created \cite{Hagemeister}. The studied skyrmions are rotationally symmetric. This suggests that their centers might relax to a lattice point which possesses the same
symmetry. A hexagonal lattice has three points showing rotational symmetry depicted in the upper inset of Fig. 3. In the following we denote these center positions as n-neighbor-centers with $n\in{2,3,6}$ according to the corresponding number of nearest-neighbor lattice points. Our LLG simulations indeed show a very good agreement with this consideration. Starting with a random configuration and using the parameters from the corresponding part of the phase diagram of Fig.1c,d, two out of three equilibrium skyrmion center configurations were obtained. The 3-neighbor-center was not observed yet. The effect that certain center positions are favored, is similar to the Peierls condition \cite{Novoselov} according to which the center of a magnetic non-collinear structure preferably occurs between the atomic sites.

The isolated skyrmion was annihilated using a magnetic field B greater than the critical field $B_c$, at which the life-times of skyrmion and ferromagnetic phase are identical (see \cite{Hagemeister}). The dynamical evolution of the skyrmion was recorded. The change in the skyrmion radius as a function of time for different field strengths is shown in the bottom panel of Fig.3. The dynamics of the skyrmion annihilation follows the same scenario for any field strength. We were able to define two dynamical stages: continuous contraction and discontinuous annihilation of isolated skyrmions. The discontinuous phase starts as soon as the skyrmion radius reaches its minimal value $R_{\rm min}(B,\frac{D}{J})$ described in Fig.2b. While the duration of the first phase decreases with increasing field strength, the second phase is practically instantaneous independently of the field strength. The inset in the bottom panel of Fig.3 shows the minimal skyrmion radius for both observed center configurations. In accordance with MC simulations of Section 3 the ultimately small skyrmion consists of four or seven atomic spins only. The value of $R_{\rm min}\approx 2.35$  $\rm {\AA}$ at zero temperature obtained in LLG simulations is close to the experimentally found skyrmion radius of 3.5 $\rm {\AA}$ at T=4.2 K \cite{Romming:PRL2015}. A slight increase of the skyrmion radius at finite temperatures is supported by the MC simulations.

The time dependent evolution of the skyrmion radius up to the first inflection point can be accurately described by an exponential fit-function
\begin{equation}\label{eq:fit1}
    R_{\frac{D}{J},R_0}(t,B)=a(B)\cdot e^{-b(B) t}+c(B),
\end{equation}
where $R_0$ is the initial skyrmion radius and the coefficients $a$, $b$ and $c$ are field-dependent functions. The coefficients of $R_{0.33,7.5{\rm\AA}}(B,t)$ for our Pd/Fe/Ir(111) based simulations correspond to:
\begin{eqnarray*}
  a(B) &\approx& -2.649{\rm\AA}+1.162B\frac{\rm\AA}{\rm T}-0.0431B^2\frac{\rm\AA}{\rm T^2} \\
  b(B) &\approx& 1.881{\rm\AA}- 0.680 B\frac{\rm\AA}{\rm T}\\
  c(B) &\approx& 9,383{\rm\AA}-1.072B\frac{\rm\AA}{\rm T}+0.0395B^2\frac{\rm\AA}{\rm T^2}.
\end{eqnarray*}

\begin{figure}
\includegraphics*[width=0.990 \columnwidth]{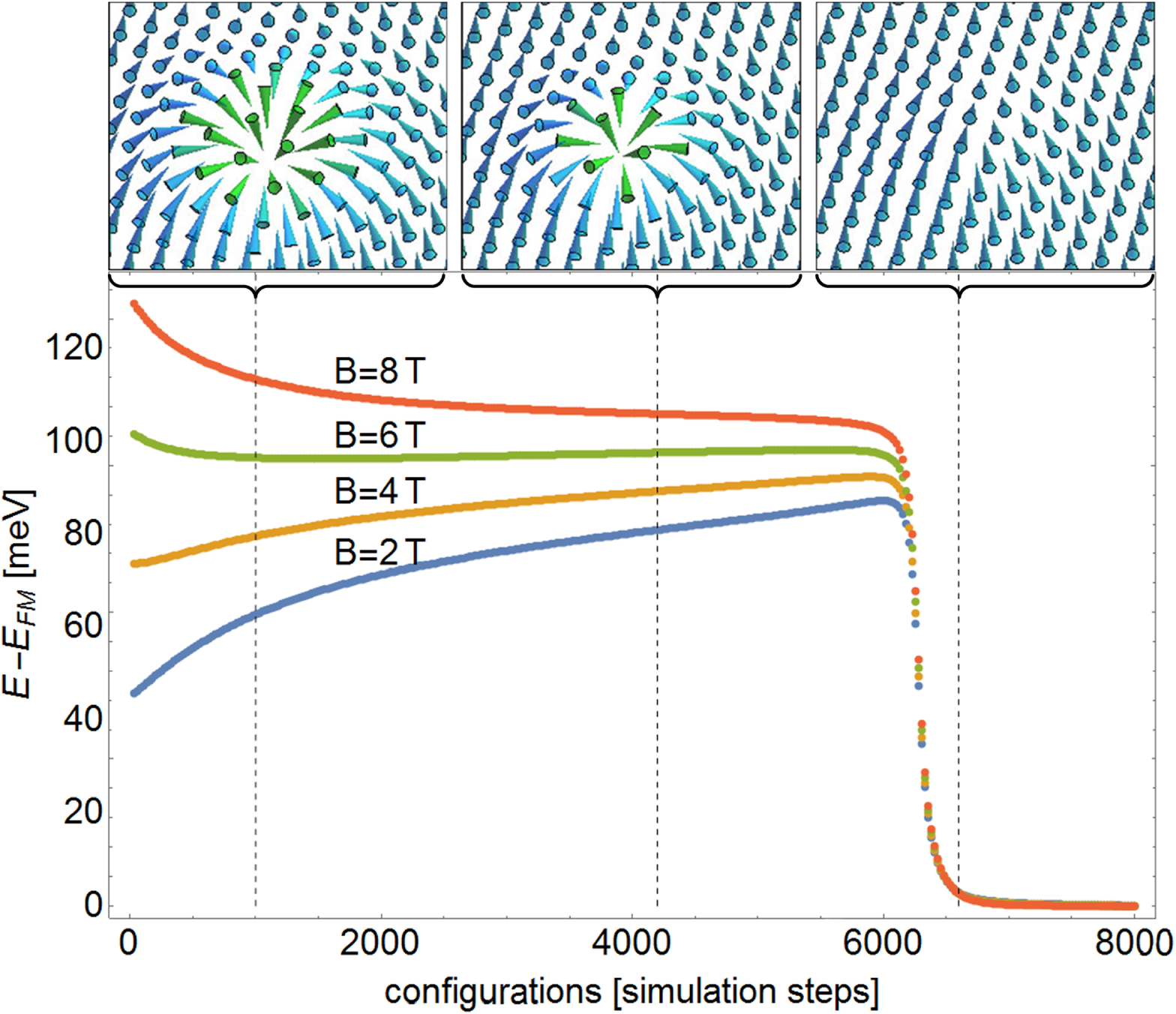}
	\caption{Statically calculated energy difference between the above given configuration and the ferromagnetic state. The sequence of configurations on the configurational axis corresponds to that obtained in deterministic LLG spin dynamics during the skyrmion annihilation in strong enough fields. Three characteristic configurations are shown in the upper panel. Fields of 8 T and 6 T are strong enough to overcome the energy barrier. For this case the configurational axis can be rescaled in time units. The lower fields would not result in skyrmion annihilation in time-dependent simulations.
	}
	\label{fig:F4}
\end{figure}
\section{Energy Barrier}

The knowledge of the strength and the spin configuration of the energy barrier between the skyrmionic and the ferromagnetic phases is extremely important for the design of skyrmion based devices. While the height of the energy barrier can be obtained from the MC simulations \cite{Hagemeister}, its micromagnetic configuration is not accessible by this technique. For that reason we address this issue here by means of the deterministic LLG spin dynamics. Our LLG simulations of the annihilation of an isolated skyrmion at different fields (see Fig.3) show that the skyrmion-to-ferromagnet transition path is identical for any $B>B_c$. The strength of magnetic field affects the time scale of the annihilation only. Hence, it is reasonable to assume that the sequence of configurations obtained in LLG calculations defines the configurational axis of the energy landscape. The role of the applied field $B$ is merely to compensate the energy barrier of the skyrmion-to-ferromagnet transition. This assumption can be crosschecked by subtraction of the energy of the ideal ferromagnetic state from the total energy of a skyrmion with particular configuration. Fig.4 gives the energy difference $E-E_{\rm FM}$ for different values of external magnetic field $B$ in configurational space, where $E$ is the energy of the corresponding micromagnetic configuration in a given applied field. Three most characteristic states on the configurational axis are also shown in this plot.

A field of 8 T is large enough to annihilate the skyrmion. Therefore, the upper curve in Fig.4 doesn't have any maximum. A skyrmion in this field has an energy, which is significantly higher than the energy barrier. When the configuration corresponding to the energy maximum on the configurational axis is reached an abrupt change in the energy can be seen. The same scenario can be observed based on the time-dependent LLG simulations: at a particular time a transition to the ferromagnetic state occurs (compare Fig.3 and the upper curve of Fig.4). The field $B\approx 6$ T is also sufficiently strong to overcome the energy barrier. However, the slope of the $E-E_{\rm FM}$ curve along the configurational axis is vanishing, indicating that the field energy is close to that of the energy barrier.
The two smaller fields of 4 T and 2 T are too weak to annihilate the skyrmion. Hence, in time-dependent LLG simulations at those field values the skyrmion would not be annihilated. However, static calculations of the energy $E-E_{\rm FM}$ in phase space show the energy maximum at the configuration corresponding to the energy barrier. As one can see in Fig.4 the steplike energy change appears always for the same configuration. This configuration matches in good approximation the observed skyrmion of minimal radius schematically depicted in Fig.3. Evidently, the energy barrier from the side of a ferromagnet slightly depends on the applied field. At small fields it is on the order of $8J$-$10 J$ in agreement with previous publications \cite{Hagemeister,Rosch}.

\section{Conclusions}

In this investigation we have obtained a comprehensive phase diagram of different skyrmionic states as a function of the characteristic density of the skyrmionic lattice. In contrast to other phase diagrams our approach permits to distinguish between the isolated skyrmions, diluted and ordered skyrmion lattices. We have shown that the field and material parameter dependence of the skyrmion size in all those phases on a discrete lattice can be well described by the continuum theory. However, we were able to find an ultimately small skyrmion of finite size, which cannot be derived from continuum theory. By analyzing the dynamics of the isolated skyrmions we were able to establish the microscopic structure of the energy barrier between the skyrmionic and the ferromagnetic states. Our results are in good agreement with previous experimental investigations of the Pd/Fe/Ir(111) systems. Therefore, we regard the results obtained by our investigation as a prerequisite for further implementation of isolated skyrmions in spintronic devices.

{\em Acknowledgements.}
Support by the DFG (SFB 668) and the EU FET Project MagicSKY (grant No. 665095) is gratefully acknowledged.


\section*{References}
\begin{thebibliography}{99}

\bibitem{Bogdanov:J.Phys.}
U. K. R\"o{\ss}ler, A. A. Leonov, and A. N. Bogdanov, J. Phys.: Conf. Ser. \textbf{303}, 012105 (2011).
\bibitem{Manton}
N. Manton and P. Sutcliffe, Topological Solitons (Cambridge University Press, 2004).
\bibitem{Roessler:Nature} R\"o{\ss}ler, A. N. Bogdanov, C. Pfleiderer, Nature (London),
\textbf{442}, 797 (2006).
\bibitem{Bogdanov89} A. N. Bogdanov and D. A. Yablonskii, Sov. Phys. JETP \textbf{68}, 101 (1989).
\bibitem{Rosza}
E. Simon, Palot\'{a}s, L. R\'{o}zsa, L. Udvardi, L. Szunyogh,  Phys. Rev. B \textbf{90}, 094410 (2014).
\bibitem{Boeni} D. Lamago, R. Georgii, C. Pfleiderer, and P. B\"oni, Physica B \textbf{385-386}, 385 (2006).
\bibitem{Tokura} X. Z. Yu, Y. Onose, N. Kanazawa, J. H. Park, J. H. Han, Y. Matsui, N. Nagaosa, and Y. Tokura, Nature (London), \textbf{465}, 901 (2010).
\bibitem{Monchesky} M. N. Wilson, A. B. Butenko, A. N. Bogdanov, and T. L. Monchesky, Phys. Rev. B \textbf{89}, 094411 (2014).
\bibitem{VonB}  K. von Bergmann, S. Heinze, M. Bode, E. Y. Vedmedenko, G. Bihlmayer, S. Bl\"ugel, and R. Wiesendanger \textbf{96}, 167203 (2006); S. Heinze et al., Nature Phys. \textbf{7}, 713 (2011).
\bibitem{Fert}
J. Sampaio, V. Cross, S. Rohart, A. Thiaville, and A. Fert, Nat. Nanotech. \textbf{8}, 839 (2013).
\bibitem{Romming:Science}
N. Romming, C. Hanneken, M. Menzel, J. E. Bickel, B. Wolter, K. von Bergmann, A. Kubetzka, and R. Wiesendanger,
Science \textbf{341}, 636 (2013).
\bibitem{Romming:PRL2015}
N. Romming, A. Kubetzka, C. Hanneken, K. von Bergmann, R. Wiesendanger, Phys. Rev. Lett. \textbf{114}, 177203 (2015).
\bibitem{Bogdanov:94}
A. Bogdanov and A. Hubert, Phys. Stat. Sol. (b) \textbf{138}, 255 (1994).
\bibitem{Butenko}
A. B. Butenko, A. A. Leonov, A. N. Bogdanov, and U. K. R\"o{\ss}ler, Phys. Rev. B \textbf{80}, 134410 (2009).
\bibitem{Vedmedenko:PRL2014}
E. Y. Vedmedenko and D. Altwein, Phys. Rev. Lett. \textbf{112}, 017206 (2014).
\bibitem{Leonov}
A. O. Leonov, T. L. Monchesky, N. Romming, A. Kubetzka, A. N. Bogdanov, and R. Wiesendanger, arXiv:1508.02155v1.
\bibitem{Banerjee:PRX2014}
S. Banerjee, J. Rowland, O. Erten, and M. Randeria, Phys. Rev. X \textbf{4}, 031045 (2014).
\bibitem{Buhrandt:PRB2013}
S. Buhrandt and L. Fritz, Phys. Rev. B 88, 195137 (2013).
\bibitem{Han}
S. D. Yi, S. Onoda, N. Nagaosa, and J. H. Han, Phys. Rev. B \textbf{80}, 2009
(2009).
\bibitem{Li}
Y.-Q. Li, Y.-H. Liu and Y. Zhou, Phys. Rev. B \textbf{84}, 205123 (2011).
\bibitem{Vedmedenko:JMMM2003}
E.Y. Vedmedenko, H. P. Oepen, and J. Kirschner, J. Magn. Magn. Mater. \textbf{256}, 237 (2003).
\bibitem{Millev:JPhysD2003}
J. Millev, E.Y. Vedmedenko, H. P. Oepen, J. Phys. D: Appl. Phys. \textbf{36}, 2945 (2003).
\bibitem{Stapelfeldt}
T. Stapelfeldt, R. Wieser, E. Y. Vedmedenko and R. Wiesendanger, Phys. Rev. Lett. \textbf{107}, 027203 (2011).
\bibitem{Hagemeister}
J. Hagemeister, N. Romming, K. von Bergmann, E.Y. Vedmedenko and R. Wiesendanger, Nature Comms. \textbf{6}, 8455 (2015).
\bibitem{Duine:PRB2015}
R. Keesman, A. O. Leonov, P. van Dieten, S. Buhrandt, G. T. Barkema, L. Fritz, and R. A. Duine, Phys. Rev. B \textbf{92}, 134405 (2015).
\bibitem{Kubetzka}
A. Kubetzka, O. Pietzsch, M. Bode, and R. Wiesendanger, Physical Review B \textbf{67}, 020401 (2003).
\bibitem{Novoselov}
K. S. Novoselov, A. K. Geim, S. V. Dubonos, E. W. Hill, and I. V. Grigorieva,
Nature \textbf{426}, 812 (2003).
\bibitem{Rosch}
C.Sch\"utte, J. Iwasaki, A. Rosch, N. Nagaosa, Phys. Rev. B {\textbf 90}, 174434 (2014).


\end{thebibliography}
\end{document}